\documentclass[twocolumn,aps,superscriptaddress]{revtex4}

\usepackage{amssymb}
\usepackage{amsmath}
\usepackage{graphicx}
\usepackage[normalem]{ulem}
\usepackage[dvips]{color}
\usepackage{appendix}

\usepackage{color}
\usepackage[normalem]{ulem}

\begin{document}

\title{Simulating fusion reactions from Coulomb explosions within a transport approach}
\author{Zhe Zhu}
\affiliation{Shanghai Institute of Applied Physics, Chinese Academy
of Sciences, Shanghai 201800, China}
\affiliation{University of Chinese Academy of Sciences, Beijing 100049, China}
\author{Jun Xu\footnote{Corresponding author: xujun@zjlab.org.cn}}
\affiliation{Shanghai Advanced Research Institute, Chinese Academy of Sciences, Shanghai 201210, China}
\affiliation{Shanghai Institute of Applied Physics, Chinese Academy of Sciences, Shanghai 201800, China}
\author{Guo-Qiang Zhang}
\affiliation{Shanghai Advanced Research Institute, Chinese Academy of Sciences, Shanghai 201210, China}
\affiliation{Shanghai Institute of Applied Physics, Chinese Academy of Sciences, Shanghai 201800, China}
\date{\today}

\begin{abstract}

We have studied nuclear fusion reactions from the Coulomb explosion of deuterium clusters induced by high-intensity laser beams within a transport approach. By incorporating the D+D $\rightarrow$ n + He$^3$ channel as inelastic collisions based on the stochastic method, we have calibrated the neutron yield from the simulation in a box system with that from the reaction rate equation. After justifying the Coulomb explosion of a single cluster by comparing results with available theoretical limits, we have then investigated the dynamics from Coulomb explosions of systems with different cluster numbers and different deuteron numbers in clusters. We find that the kinetic energy spectrum of deuterons at the final stage is different from that when neutrons are abundantly produced, corresponding to significantly different reaction rates. We also extrapolate the neutron yield result from small systems to large systems based on an intuitive parameterized form and compare with the available experimental result. The present framework can be extended by incorporating more channels, and useful for further studies of nuclear fusion reactions in plasma systems at higher energies reached in more recent experiments.

\end{abstract}

\maketitle

\section{INTRODUCTION}
\label{introduction}

Nuclear fusion reactions on cluster targets induced by high-intensity laser beams have been an active research field for two decades~\cite{Dit98,Smi00}. Experimentally, cryogenically cooled deuterium cluster targets or deuterated methane cluster targets at around room temperature are irradiated by high-intensity femtosecond laser to drive fusion reactions. It has been shown that this process can be well described by the Coulomb explosion model~\cite{Kra02}, where almost all electrons are stripped off by the laser and removed from clusters in a short time, during which the ions can be considered nearly stationary. What remain in cluster are ions in the liquid density, which then explode by Coulomb repulsion and are accelerated to exceeding keV energy, making fusion reactions become possible.

Fusion reactions on clusters induced by laser first attracted people's attention because it has the potential to become a stable new source of neutrons, which may have potentially wide applications in other research fields, e.g., material science, etc. As shown in the pioneer study in Ref.~\cite{Dit99}, a laser pulse generated by a desktop laser based on the chirped pulse amplification technology can be used to obtain neutrons by irradiating on deuterium clusters, with an efficiency of about $10^5$ fusion neutrons per joule of the incident laser energy. Later on, attempts were made to increase neutron yield by investigating target clusters of different compositions (see, e.g., Ref.~\cite{Gri02}). The details of the fusion process, such as the relative contribution to the fusion yield from both beam target and intrafilament fusions, were also investigated~\cite{Mad04}. Besides studies on the mechanism of fusions, the reaction products can also be used as a probe of properties of the plasma produced by the irritation of high-intensity laser beams, e.g., the ratio of two different fusion products, i.e., neutrons and protons from D(d,$^3$He)n and $^3$He(d,p)$^4$He, can be used to probe the temperature of the plasma~\cite{Ban13}. In addition, such fusion reactions can also be used to study those inside stellars or during the early evolution of the Universe. Particularly, the products of the fusion reactions were used to measure the $S$ factor of the $^3$He(d,p)$^4$He reaction~\cite{Bar13} at rather low center-of-mass (C.M.) energies, an important quantity to understand nucleosynthesis.

Since fusion reactions induced by laser beams have been a hot topic as discussed above, a theoretical model is called for to understand the dynamics of nuclear reactions in the plasma. In the studies of Refs.~\cite{Ban13,Bar13}, it is assumed that the ions are in thermal equilibrium and the reaction rate is calculated by assuming that the particle velocities follow a Maxwell-Boltzmann (MB) distribution. However, the life time of the plasma is much shorter compared to the relaxation time for thermal equilibrium under short-range Coulomb collisions~\cite{Kra02,Spi67,Smi01}. In the picture of Coulomb explosion, the energy distribution of ions in the plasma is mostly driven by the electrostatic field and is related to the size distribution of clusters~\cite{Kra02,Dit99,Zwe02}. In such non-equilibrium situation, it is proper to study the dynamics of the plasma with transport simulations. In the present framework, we employ the EPOCH model, a typical particle-in-cell (PIC) transport approach, to simulate the Coulomb explosion of clusters with electrons completely stripped by laser beams. The nuclear fusion reactions are incorporated by introducing inelastic scattering channels between ions with the stochastic method. In the simulation of Coulomb explosion, we set the deuteron number density inside clusters to be $4.9 \times 10^{10} $ $\mu$m$^{-3}$, the average deuteron number in each cluster to be about 10000, and the distance between deuterium clusters about 0.056 $\mu$m, which can be achieved by the experiments as in Refs.~\cite{Dit99,Zha17}. We find that the kinetic energy spectrum of ions at the final stage of the reaction, which depends on the cluster properties, is different from that when fusion reactions actively occur, corresponding to significantly different reaction rates. We also find ways to extrapolate the neutron yield result from small systems to large systems.

The rest part of the paper is organized as follows. Section~\ref{theory} gives the theoretical framework of the present study, with a brief overview of the EPOCH model and the description of method to incorporate nuclear fusion reactions. Section~\ref{results} calibrates the neutron yield from simulations in a box system by comparing results from the reaction rate equation, compares simulation results of the Coulomb explosion of a single cluster with theoretical limits, and then discusses simulation results from Coulomb explosions of many deuterium clusters in detail. We conclude and outlook in Sec.~\ref{summary}.

\section{Theoretical framework}
\label{theory}

The PIC approach has been a tool widely used in the simulation of plasma physics since the 1970s. In recent years, the PIC code has been continuously developed to include effects such as collision, ionization, QED, etc. In the following, we briefly remind the reader about the main features of the EPOCH model, and mainly focus on how we incorporate the inelastic nuclear reaction channels. For details of the EPOCH code, we refer the reader to Ref.~\cite{Arb15}.

\subsection{Framework of EPOCH}
\label{framework}

The dynamics in the EPOCH framework mainly contains two parts, i.e., the collisionless part and the collision part. The collisionless part consists of the propagation of charged particles under the electromagnetic (EM) field~\cite{Bor70}, which leads to electric currents~\cite{Esi01,Vil92}, and the calculation of the EM field by solving Maxwell's equations based on the current generated by the motions of charged particles on a fixed spatial grid~\cite{Yee66}. The particles in EPOCH only couple to the EM field via current deposition, which ensures that $\nabla \cdot \vec{E}= \rho_q/\epsilon_0$ is always satisfied, with $\vec{E}$, $\rho_q$, and $\epsilon_0$ being respectively the electric field, the charge density, and the dielectric constant, if the initial conditions of the system are consistent with Gauss's law. Each simulation particle in EPOCH represents a certain number of real particles, with this number called the weight. In the present study, we set the weight of all deuterons to be 1 in order to describe properly the Coulomb explosion dynamics of clusters containing many deuterons, while those of produced particles from fusion reactions are much smaller than 1. Each simulation particle has a finite size with its spatial distribution described by the shape function, which can be adjusted artificially according to the problem to be addressed. In the present study, a shape function with a 5th-order B-spline method is used to avoid self-heating~\cite{Arb15}. The relevant collision part in the present study is the Coulomb collision based on the approach by Sentoku and Kemp~\cite{Sen98}, where a particle can only collide with another particle in the same cell, and the collision algorithm is executed in each cell in the simulation area. This approach treats short-range Coulomb collisions stochastically in momentum space, with the energy conserved perfectly in each collision but momentum conserved on average. For a recent improved treatment on Coulomb collisions, see Ref.~\cite{Hig20}. 

\subsection{Incorporating nuclear reaction channel}

We incorporate the inelastic D+D $\rightarrow$ n+$^3$He channel based on the stochastic method commonly used in simulations of heavy-ion collisions, where the probability of a $2\rightarrow2$ reaction in a time interval $\Delta t$ and box volume $(\Delta x)^3$ is~\cite{Xu05}
\begin{equation}\label{Psto}
P = v_{mol} \sigma \frac{\Delta t}{(\Delta x)^3}
\end{equation}
with
\begin{equation}
v_{mol} = \frac{c\sqrt{[s-(M_1+M_2)^2c^4][s-(M_1-M_2)^2c^4]}}{2E_1E_2}
\end{equation}
being the M{\o}ller velocity in SI unit, where $s$ is the square of the C.M. energy, $M_{1(2)}$ and $E_{1(2)}$ are the mass and energy of particle 1(2), and $\sigma$ is the cross section of D+D $\rightarrow$ n+$^3$He reaction. In the nonrelativistic limit, the M{\o}ller velocity reduces to the relative velocity $|\vec{v}_1-\vec{v}_2|$, with $\vec{v}_{1(2)}$ being the velocity of particle 1(2). According to Eq.~(\ref{Psto}), the collision number per unit time per unit volume is exactly the reaction rate $\langle v_{mol} \sigma \rangle$, with $\langle ... \rangle$ being the average in local phase space. In principle, $\Delta t$ and $(\Delta x)^3$ are required to be as small as possible but still contain enough stimulation particles, so that the Boltzmann limit of the collision rate can be achieved~\cite{Bab89}. We use $\Delta t = 0.002$ fs and $(\Delta x)^3=0.01\times 0.01 \times 0.01$ $\mu$m$^3$ for evaluating the reaction probability in simulating the dynamics of Coulomb explosions. The momenta of produced particles, i.e., n and $^3$He, are sampled isotropically in the C.M. frame of inelastic D+D $\rightarrow$ n+$^3$He collisions, and then Lorentz-boosted to the calculational frame. Both momentum and energy are conserved in the initial and final state of D+D $\rightarrow$ n+$^3$He collisions, according to the treatment of inelastic collisions in the appendix B of Ref.~\cite{Ber88}.

Equation~(\ref{Psto}) is valid for collisions between microscopic particles with weight equal to 1 as in the present study, and can be generalized to incorporate other channels and weight corrections. For a more general case, the collision probability between particles with weight $w_1$ and $w_2$ is~\cite{Hig19}
\begin{equation}
P = w_1 w_2 v_{mol} \sigma \frac{\Delta t}{(\Delta x)^3}.
\end{equation}
The number of collisions per unit time per unit volume in the system containing particle species 1 and 2 with their number densities respectively $\rho_1$ and $\rho_2$ can then be expressed as~\cite{ll}
\begin{equation}
\frac{d N_{coll}}{d t dV} = \frac{1}{1+g_{12}} w_1 w_2 \rho_1 \rho_2 v_{mol} \sigma,
\end{equation}
with $g_{12}=1$ for identical particles and $g_{12}=0$ otherwise.

\begin{figure}[ht]
\includegraphics[scale=0.3]{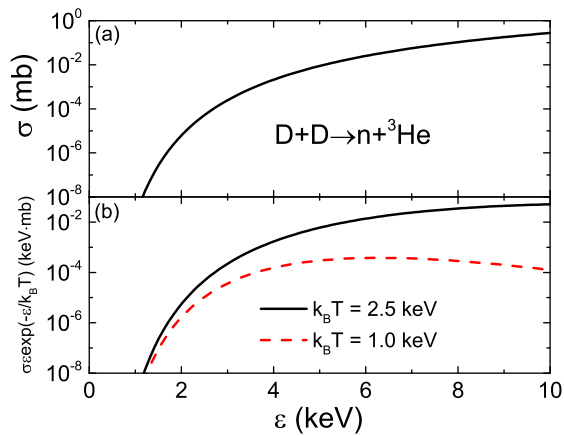}
\caption{(Color online) Cross section of D+D $\rightarrow$ n+$^3$He (upper) and the integrated part of the reaction rate in Eq.~(\ref{rate}) at different temperatures (lower) as a function of the total kinetic energy of deuterons in their C.M. frame.}\label{fig1}
\end{figure}

In the present study, we adopt the kinetic energy dependence of $\sigma$ in Eq.~(\ref{Psto}) as~\cite{Bos92}
\begin{equation}
\sigma = \frac{S(\epsilon)}{\epsilon \cdot \exp(B_G/\sqrt{\epsilon})},
\end{equation}
where $B_G$ is the Gamov constant taken as 31.3970 $\sqrt{\text{keV}}$ for D+D reactions, $\epsilon$ is the total kinetic energy of two deuterons in their C.M. frame, and $S(\epsilon)$ is the S factor parameterized as~\cite{Bos92}
\begin{eqnarray}
S(\epsilon) &=& 5.3701\times10^4 + 3.3027\times10^2 \epsilon - 0.12706 \epsilon^2 \notag\\
&+& 2.9327\times10^{-5} \epsilon^3 - 2.5151\times10^{-9} \epsilon^4,
\end{eqnarray}
with $\epsilon$ in keV and $S$ in keV $\cdot$ mb. The dependence of $\sigma$ on $\epsilon$ is displayed in Fig.~\ref{fig1} (a), where $\sigma$ is seen to increase exponentially with increasing $\epsilon$.

The probability of D+D fusions is actually very low in the situation considered here. As shown in Ref.~\cite{Bos92}, in the thermal equilibrium condition with the temperature of deuterons about 50 keV, the reaction rate $\langle v_{mol} \sigma \rangle$ of D+D $\rightarrow$ n+$^3$He is about $10^{-20}$ $\mu$m$^3$ fs$^{-1}$, which is even higher than that in the present study. With the probability of inelastic collisions given by Eq.~(\ref{Psto}), it is possible to control the number of simulation particles in the final state, i.e., n and $^3$He, by choosing a proper weight for them. Equivalently, we modify the probability of inelastic collisions to $P/P_0$ with $P_0=8.115 \times 10^{-16}$ larger than the maximum value of all possible $P$, so that the weight of the produced particles (n or $^3$He) is $P_0$. In principle, we should also incorporate the inverse channel, i.e., n+$^3$He $\rightarrow$ D+D, with the corresponding cross section determined by the detailed balance condition. However, since the reaction rate is very small and the numbers of n and $^3$He are much smaller than that of D, it is safe to neglect the inverse channel in the simulation. The nuclear fusion reactions are just perturbation to the dynamics of deuterons in the present study, different from the situation in Ref.~\cite{Hig19}.

\section{Results and discussions}
\label{results}

In order to validate that we incorporate the nuclear fusion reaction correctly, we first do simulation in a box system with the periodic boundary condition, where the results can be compared with the theoretical limit provided by the time integral of the reaction rate. Next, we compare the dynamics of the Coulomb explosion of a single deuterium cluster from the EPOCH simulation with the theoretical limit. With the well calibrated nuclear reaction treatment and the dynamics of deuterons, we then simulate Coulomb explosions of many deuterium clusters with the free boundary condition, i.e., particles are allowed to escape from the simulation region, and study the production of neutrons in systems with different numbers of clusters, in order to investigate the finite-size effect and extrapolate the results to macroscopic systems. Typically, we try to understand the relation between the kinetic energy spectrum of deuterons and the production of neutrons. In the simulation of the Coulomb explosion, we have fixed the average space distance between clusters and the deuteron density inside clusters, while the average number as well as the number distribution of deuterons in clusters are varied, in order to vary the kinetic energy spectra of deuterons and discuss the effect on the production power of neutrons. In both simulations of the box system and the Coulomb explosion, we take values of parameters by referencing the experimental conditions in Refs.~\cite{Dit99,Zha17}.

\subsection{Fusion reaction in a box system}

\begin{figure}[ht]
	\includegraphics[scale=0.35]{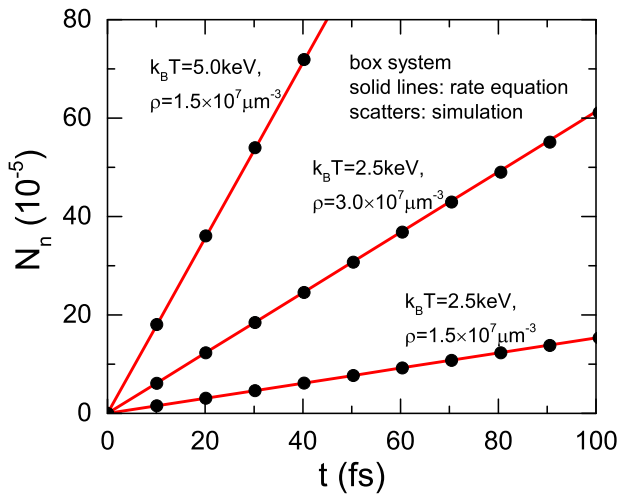}
	\caption{(Color online) Neutron number as a function of time from the rate equation [Eq.~(\ref{rateeq})] and from the simulation with the stochastic method in a box system at different deuteron densities $\rho$ and temperatures $T$.}\label{fig2}
\end{figure}

The simulation is carried out in a box with the volume $V=2\times2\times2$ $\mu$m$^3$, the time scale about 100 fs, and at typical deuteron densities $\rho$ and temperatures $T$ when neutrons are expected to be abundantly produced during the Coulomb explosion. The box is divided into $20 \times 20 \times 20$ cells, with the grid length about 0.1 $\mu$m, same for evaluating short-range Coulomb collisions and inelastic collisions. Since the box system is uniform, the EM field doesn't play a role. The velocities of deuterons are prepared in the MB distribution, i.e., $f(v) = \left(\frac{m_D}{2\pi k_BT}\right)^{3/2} \exp(-\frac{m_Dv^2}{2k_BT})$, with $m_D$ being the deuteron mass, and the distribution is maintained in the presence of short-range Coulomb collisions. Thus the neutron yield from D+D $\rightarrow$ n+$^3$He reactions can be calculated through
\begin{equation} \label{rateeq}
N_n=\frac{1}{2}V\rho^2\int \langle \sigma v_{mol} \rangle dt,
\end{equation}
where
\begin{eqnarray} \label{rate}
\langle \sigma v_{mol} \rangle &=& \int d^3 v_1 d^3 v_2 \sigma (\vec{v}_1,\vec{v}_2) |\vec{v}_1-\vec{v}_2| f(v_1) f(v_2) \notag\\
&=& \frac{4}{(2\pi m_D)^{1/2}}\frac{1}{(k_B T)^{3/2}} \int_0^\infty \sigma (\epsilon) \epsilon \exp\left(-\frac{\epsilon}{k_B T}\right) d \epsilon \notag\\
\end{eqnarray}
is the thermal averaged reaction rate, with the integral over the total kinetic energy $\epsilon$ of two deuterons in their C.M. frame, and the $\epsilon$ dependence of the integrated function is shown in Fig.~\ref{fig1} (b), with the Boltzmann factors for different temperatures selecting the contribution of $\sigma$ at different $\epsilon$ regions. Neutron yields from both the reaction rate equation and box simulation using the stochastic method at different densities and temperatures are compared in Fig.~\ref{fig2}, where the time scale of about 100 fs is chosen to be the typical one for the duration of the Coulomb explosion and nuclear reaction depending on the cluster spacing used in the present study, to be discussed later. It is seen that in the density and temperature ranges considered here, the linearly increasing neutron numbers from the two methods are almost on the top of each other, justifying the validity and accuracy of the stochastic method used for incorporating fusion reactions.

\subsection{Coulomb explosion of a single cluster}

Before we simulate the fusion reactions from the deuteron-deuteron collisions caused by Coulomb explosions of many deuterium clusters, we first simulate the Coulomb explosion of a single deuterium cluster, and compare the results with available theoretical limits. The initial coordinates of deuterons are uniformly generated within a sphere in each cluster, while their initial momenta are negligibly small at room temperature. According to the assumption of the pure Coulomb explosion model, all electrons in deuterium clusters are stripped off in one laser pulse duration. The local number density of deuterium clusters is set to be $\rho_0=4.9 \times 10^{10} $ $\mu$m$^{-3}$~\cite{Zha17}. The numbers of deuterium atoms in each cluster can vary from a few hundred to a few ten thousand, and the radius of the cluster can vary from $10^{-3}$ to $10^{-2}$ $\mu$m. In the default case, we set the deuteron number of $N_c=11392$ in each cluster particle in this simulation, so the radius of each cluster is about $R_0=0.0038$ $\mu$m. The gird size for the calculation of the EM field and short-range Coulomb collisions is about $0.001$ $\mu$m, so the radius of clusters is about 4 times the grid size. We calculate the initial electrostatic field distribution through Gauss's law as the input for EPOCH, to ensure that the motion of particles and the evolution of electromagnetic fields are simulated in a consistent way.

\begin{figure}[h]
	\includegraphics[scale=0.35]{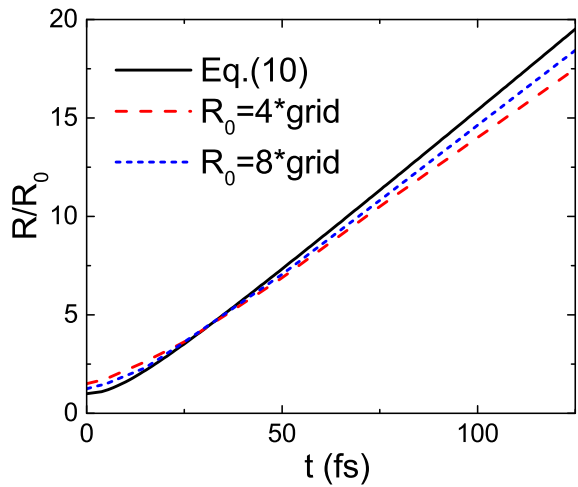}
	\caption{(Color online) Comparison of the time evolution of the explosion distance from EPOCH simulations with different grid sizes to that from Eq.~(\ref{Rt}).}\label{fig2.5}
\end{figure}

According to the energy conservation law, the distance $R$ between the explosion surface and the center of the cluster evolves with time $t$ according to the relation~\cite{Kra02}
\begin{equation}\label{ec}
\frac{1}{2} m_D \left (  \frac{dR}{dt} \right)^2 = \frac{N_ce^2}{4\pi\epsilon_0}\left( \frac{1}{R_0} - \frac{1}{R}\right).
\end{equation}
The time evolution of $R$ can be obtained by integrating the above equation, i.e.,
\begin{eqnarray}\label{Rt}
t &=& \sqrt{\frac{\pi \epsilon_0 m_D R_0}{2N_c e^2}} \notag\\
&\times&\left[2\sqrt{R(R-R_0)}+R_0\ln \left(\frac{2R-R_0+2\sqrt{R(R-R0)}} {R_0}\right) \right]. \notag\\
\end{eqnarray}
Figure~\ref{fig2.5} compares the time evolution of the explosion distance $R$ from Eq.~(\ref{Rt}) and from EPOCH simulations with different grid sizes for the calculation of the electrostatic field. The slightly different initial $R$ values from different grid sizes are due to the surface smearing in sampling deuterons coordinates within a sphere by EPOCH. One sees that results from transport simulations agree with that from Eq.~(\ref{Rt}) reasonably well. The difference between the simulation result and that from Eq.~(\ref{Rt}) can always be reduced with a smaller grid size.

\begin{figure}[h]
	\includegraphics[scale=0.5]{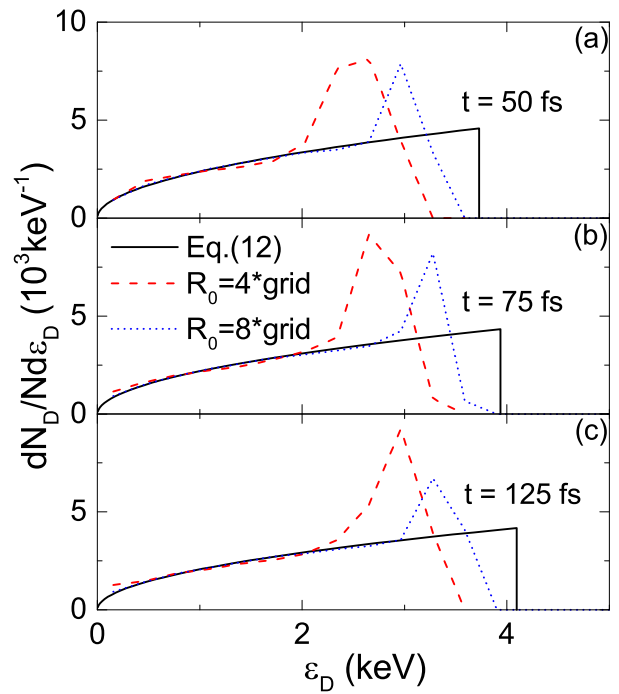}
	\caption{(Color online) Comparison of the deuteron kinetic energy spectra at $t=50$, 75, and 125 fs from EPOCH simulations with different grid sizes to those from Eq.~(\ref{de}).}\label{fig2.8}
\end{figure}

For a single spherical charged cluster, the kinetic energy spectrum of deuterons has an analytical solution, and at $t \rightarrow \infty$ the number density of deuterons at a particular kinetic energy $\epsilon_D$ is proportional to $\sqrt{\epsilon_D}$~\cite{Zwe00,Zwe02}. Following Eq.~(\ref{ec}), the kinetic energy for a deuteron at the initial radius $r$ ($0<r<R_0$) is
\begin{equation}\label{ed}
\epsilon_D = \frac{\rho_0e^2}{3\epsilon_0} \left(r^2-\frac{r^3}{r^\prime}\right),
\end{equation}
where $r^\prime$ is the explosion distance at time $t$ for the shell with the initial radius $r$. For $r=R_0$ and $r^\prime=R$, Equation~(\ref{ed}) reduces to Eq.~(\ref{ec}). By using the deuteron number $dN=4\pi r^2\rho_0 dr$ inside the shell of initial radius $r$ and thickness $dr$, and taking the derivation of Eq.~(\ref{ed}) with respective to $dr$, we can get the following relation
\begin{equation}\label{de}
\frac{dN}{d\epsilon_D} = \frac{12\pi \epsilon_0 r}{e^2 \left( 2 - 3\frac{r}{r^\prime} + \frac{r^2}{{r^\prime}^2} \frac{dr^\prime}{dr} \right)}.
\end{equation}
The analytic solution of the kinetic energy spectrum from the Coulomb explosion of a single cluster at arbitrary time can be obtained by combining the above expression with Eq.~(\ref{ed}) and a similar $r^\prime(r,t)$ relation as Eq.~(\ref{Rt}). Since the high-energy part is contributed from deuterons initially inside a shell with a larger $r$, these deuterons expand and are accelerated with time. Figure~\ref{fig2.8} compares the deuteron kinetic energy spectra at different times from Eq.~(\ref{de}) and those from EPOCH simulations with different grid sizes for the calculation of the electrostatic field. One sees that the EPOCH simulation reproduces well the kinetic energy spectra at low-energy part but not at high-energy part, where the agreement can always be improved by using a smaller grid size. This is due to the sharp density distribution on the cluster surface. In the more realistic case, the density density is diffusive on the surface of clusters, and the agreement will be better.

\subsection{Coulomb explosion of many clusters}

In the simulation of many deuterium clusters, we set the distance between neighboring cluster about 15 times the size of the cluster, i.e., $\sim 0.056$ $\mu$m, so that the density after the Coulomb explosion is reduced by $10^3-10^4$ compared to the initial density of clusters. Due to the lacking of computational power, we simulate systems much smaller than that in real experiments, while possible extrapolations to large systems are discussed. Typically, systems consisting of 8, 27, and 64 clusters are simulated, with the simulation area set to be $260 \times 260 \times 260$ cells, $320 \times 320 \times 320$ cells, and $380 \times 380 \times 380$ cells, respectively for each scenario. The average number of simulation particles in each cell is about 0.013, and this means that inelastic collisions are unlikely to take place, compared to the short-range Coulomb collisions. Therefore, almost 1000 adjacent cells are formed into a big cell artificially for the stochastic method of inelastic collisions to be executed.

\begin{figure*}[ht]
	\includegraphics[scale=0.25]{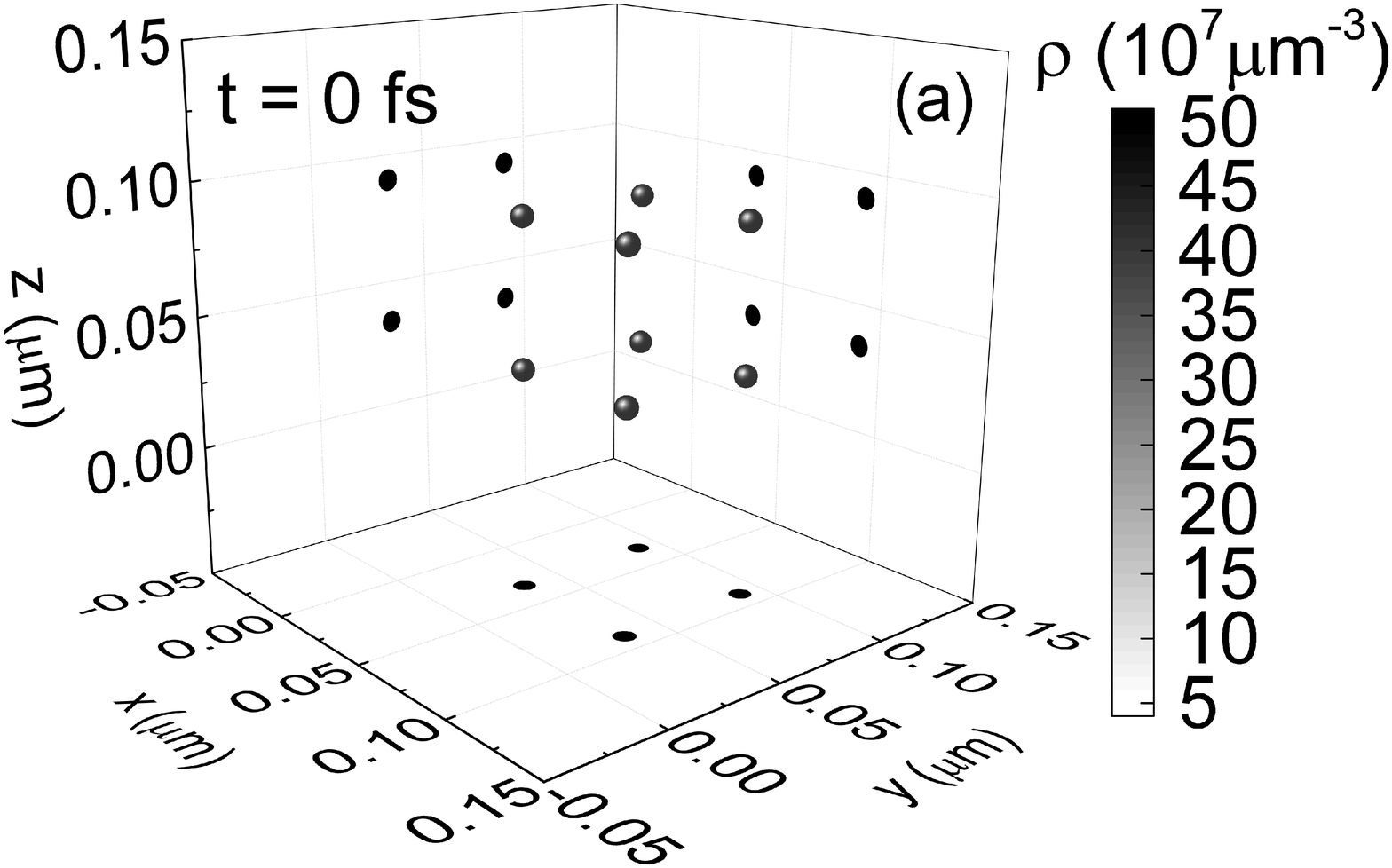} \includegraphics[scale=0.25]{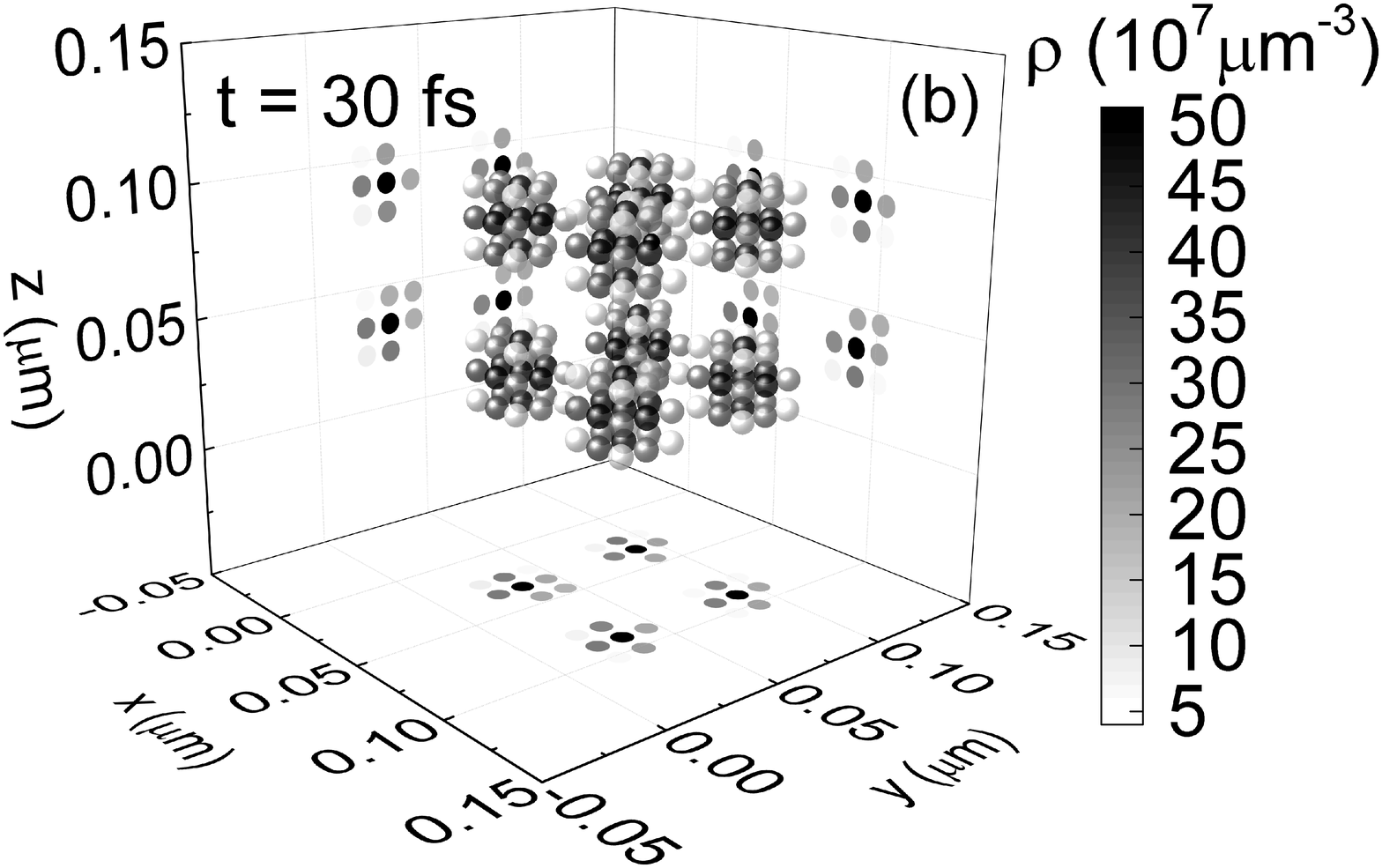} \\
\vspace{3mm}
	\includegraphics[scale=0.25]{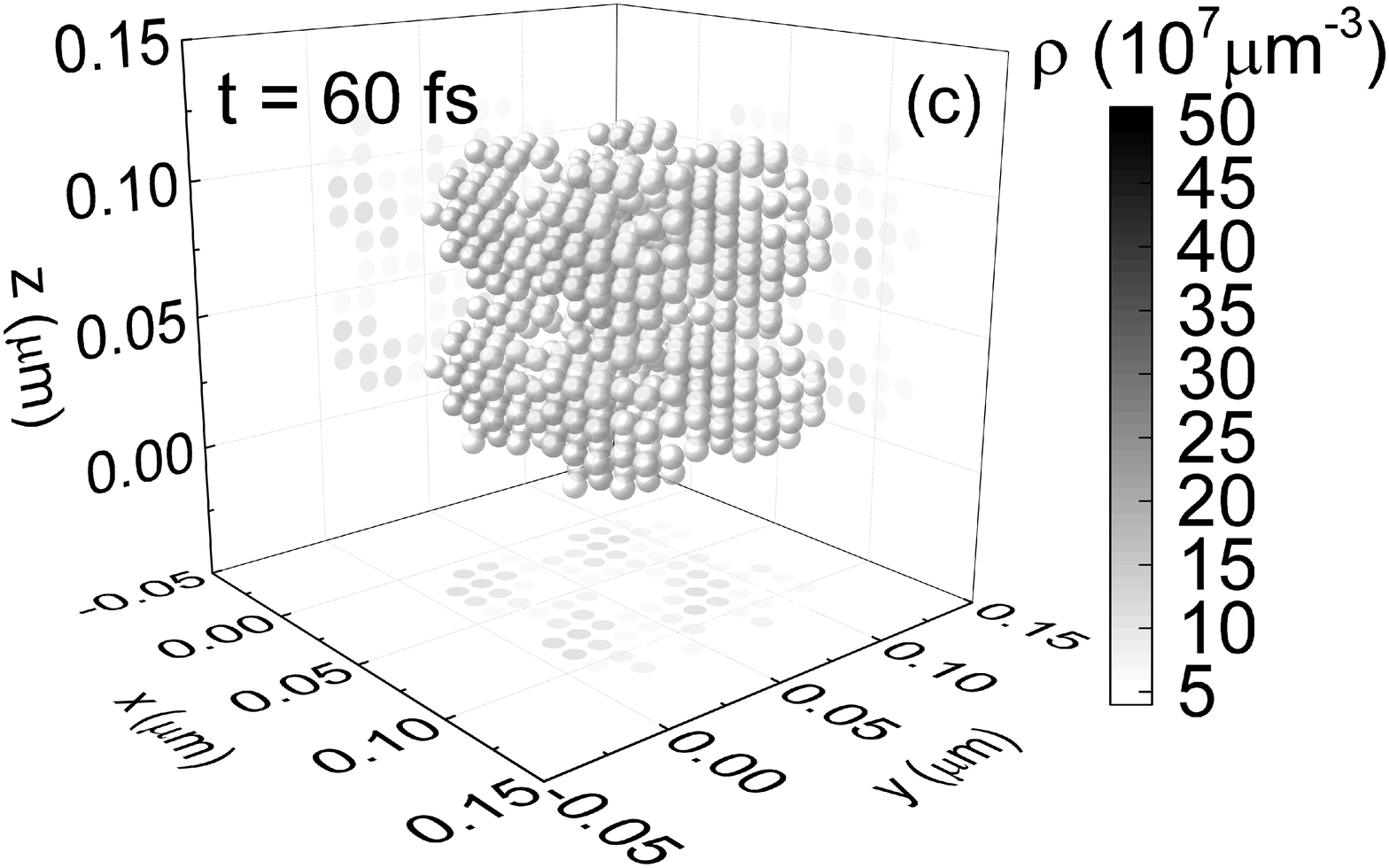} \includegraphics[scale=0.25]{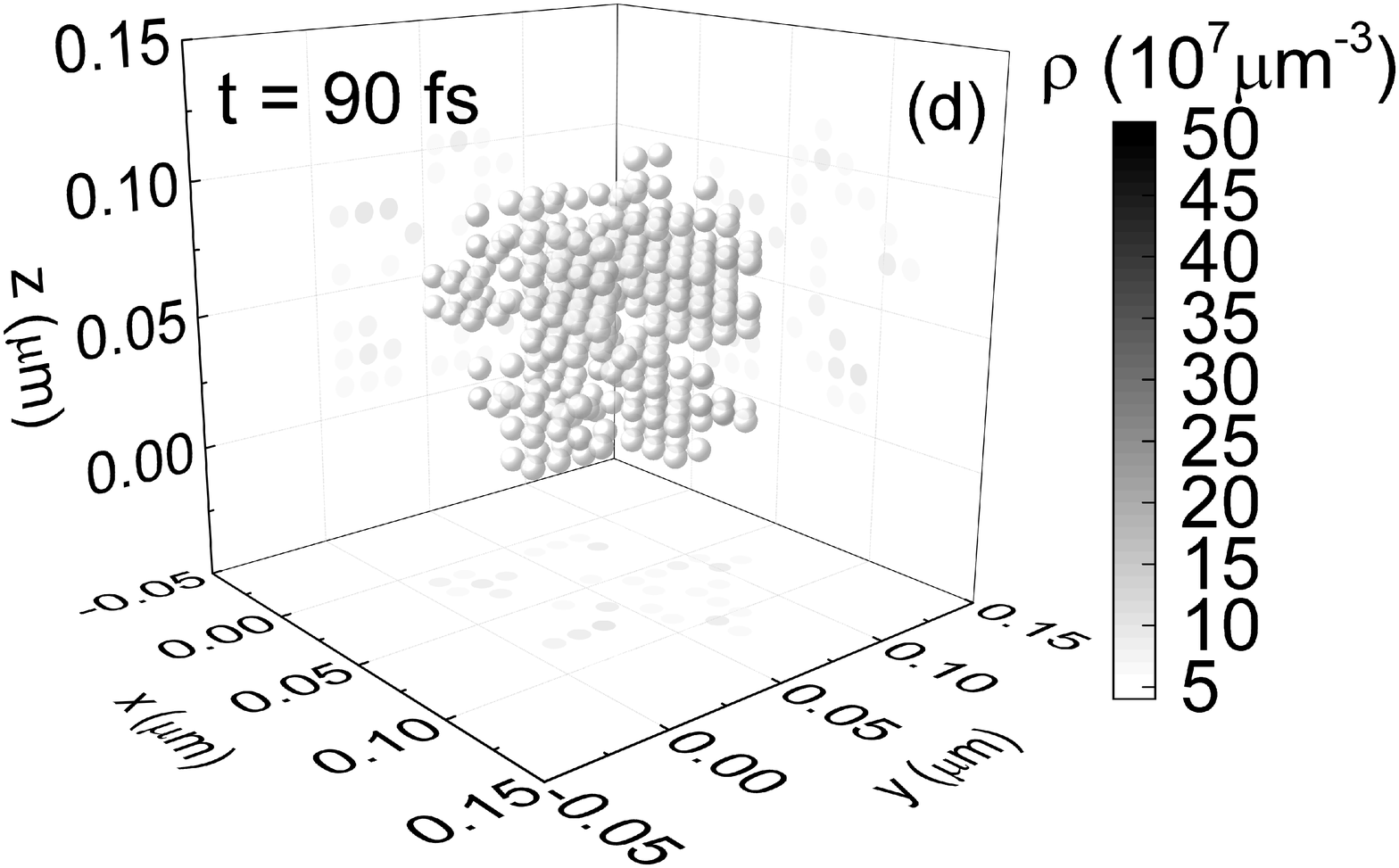}
	\caption{Density distributions from the Coulomb explosion of 8 deuterium clusters at $t=0$ (a), 30 (b), 60 (c), and 90 (d) fs. Balls with colors from black to white represent high to low deuteron densities in local cells as shown by the color scale in each panel, while cells with very low densities ($<5\times10^7 \mu$m$^{-3}$) are not displayed.}\label{fig3}
\end{figure*}

Figure~\ref{fig3} displays the intuitive picture of the density evolution from the Coulomb explosion of 8 deuterium clusters initially located at regular lattice positions. One sees that the initial size of clusters is much smaller than the system size. At $t=30$ fs the clusters expand to a much larger size, while collisions between deuterons from different clusters, which is the main source of neutron production, are still rare. At $t=60$ fs, overlaps from the expansion of different clusters are observed and extensive collisions are expected to happen. The density further drops at $t=90$ fs. The above observations are of course consistent with the time evolution of the surface explosion distance in Fig.~\ref{fig2.5}, from which one expects significant interaction between neighboring cluster begins around $t=50$ fs for $R \approx 8 R_0$, while the time scale of the whole interaction between neighboring cluster is about $t = 100$ fs for $R \approx 15 R_0$, which is the same time scale for box calculation as shown in Fig.~\ref{fig2}. Later on the system becomes dilute and many deuterons escape from the system of a limited simulation size with the free boundary condition. The density of these deuterons is very dilute, and they move in the same expanding direction, so they generally do not collide and thus do not contribute to the neutron production. This surface effect is expected to be less important with the increasing size of the system. We note that Fig.~\ref{fig3} serves as an illustration with clusters initiated at regular lattice positions, while in real simulations the initial positions of clusters are randomized in the system with the constraint that they do not overlap.

\begin{figure}[h]
	\includegraphics[scale=0.35]{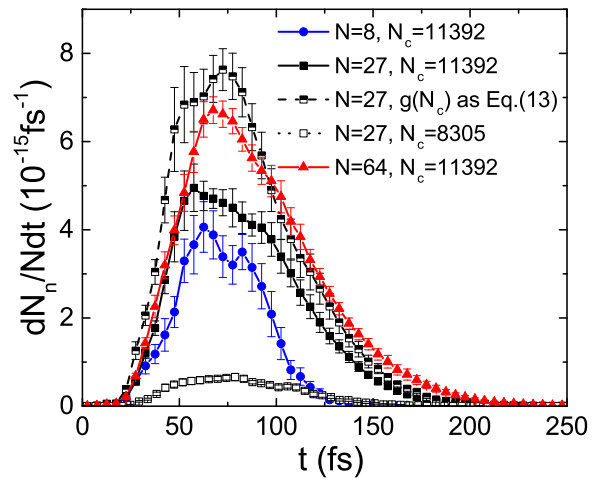}
	\caption{(Color online) Average neutron production rate per cluster $dN_n/Ndt$ as a function of time from simulating Coulomb explosions of $N=8$, 27, and 64 deuterium clusters and with different deuteron numbers $N_c$ in each cluster.}\label{fig4}
\end{figure}

Figure~\ref{fig4} displays the average neutron production rate per cluster as a function of time from the Coulomb explosions of 8, 27, and 64 deuterium clusters. In order to get better statistics, we have generated 10 simulation events for each scenario. It is seen that the neutron production rate peaks at around 75 fs. The neutron production rate per cluster increases with the increasing system size, as a result of more collisions between deuterons from the Coulomb explosions of different clusters, while the surface deuterons that first leave the simulation region do not contribute to nuclear reactions. The increasing trend reduces with the increasing size as the surface effect becomes less important for large systems. The result from 27 clusters but with a smaller $N_c=8305$ is also compared. If the energy spectrum of deuterons are similar, it is expected that the neutron yield is proportional to the collision number and thus $N_c^2$. However, the neutron yield after integrating the production rate is much smaller with a smaller $N_c$, i.e., $(8305/11392)^2 \approx 0.53$ compared to about $1/10$ from Fig.~\ref{fig4}. This is due to the different energy spectra of deuterons from different $N_c$, to be shown later. We have also compared results from 27 clusters but with $N_c$ following a log-normal distribution, i.e.,
\begin{equation}\label{ln}
g(N_c) = \frac{1}{\sqrt{2\pi} a N_c} \exp \left[ -\frac{1}{2a^2} (\ln N_c - b)^2 \right],
\end{equation}
with $a=0.231$ and $b=9.314$. In this way, the average $N_c$ is 11392, the same as in the default calculation, while the standard deviation is 2667. Even with the same total deuteron number, it is seen from Fig.~\ref{fig4} that the neutron yield from the case of a log-normal distribution with the same average $N_c$ is much larger than that from the simulation with equal $N_c$ for all clusters. This is again understandable since the released energy from a Coulomb explosion of a deuterium cluster is not proportional to $N_c$ but to $N_c^{5/3}$~\cite{Zha17}. Thus, clustering systems with $N_c$ in the log-normal distribution on average lead to more energetic deuterons and thus larger neutron yield, compared to systems with a fixed $N_c$, due to the non-linear relation between the energy release and $N_c$. The dependence of neutron yield on $N_c$ has been observed experimentally (see, e.g., Ref.~\cite{Zwe00}).

\begin{figure}[ht]
	\includegraphics[scale=0.5]{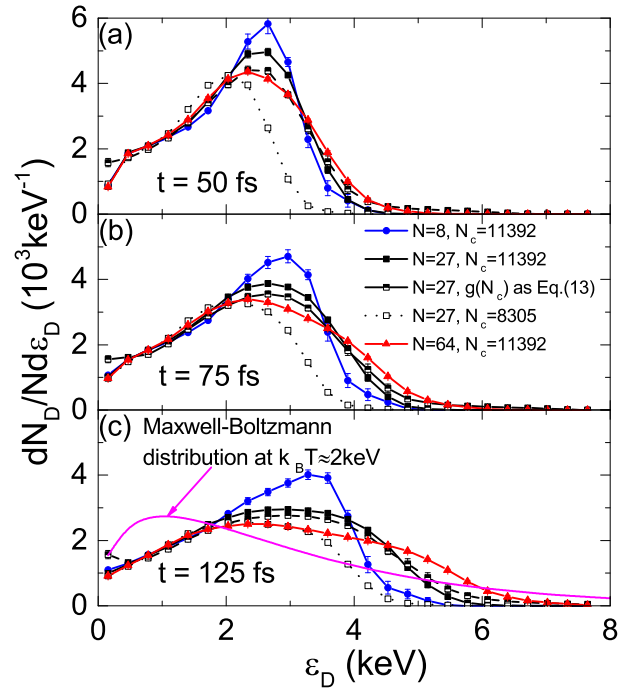}
	\caption{(Color online) Average kinetic energy spectrum of deuterons per cluster at different times from simulating Coulomb explosions of $N=8$, 27, and 64 deuterium clusters and with different deuteron numbers $N_c$ in each cluster. }\label{fig5}
\end{figure}

Figure~\ref{fig5} displays the average kinetic energy spectrum of deuterons per cluster at different times in different scenarios. Due to the Coulomb potential between different clusters, the kinetic energy spectrum becomes broader compared to that from the Coulomb explosion of a single cluster, and the stiffness of the spectrum increases with the increasing number of clusters. One sees that the distribution becomes broader with more clusters and at later times. The broadening of the kinetic energy spectrum is due to the long-range repulsive interaction from the electrostatic field rather than due to short-range Coulomb collisions, with the latter having a much longer relaxation time compared with the life time of the system considered here~\cite{Kra02,Spi67,Smi01}. The long-range Coulomb potential is expected to further stiffen the kinetic energy spectrum of deuterons at later times, while after $t=125$ fs some of deuterons escape from the system and are thus not countable or not further accelerated by the electric field in the simulation. With a smaller $N_c$, the energy release from the Coulomb explosion is much weaker, leading to a softer energy spectrum, and thus a weaker production power of neutrons. The opposite is observed for the case with a log-normal distribution for $N_c$, as already discussed above. The MB distribution with the same total deuteron number $N_D$ and the same total energy as in the case of 64 clusters are plotted in Fig.~\ref{fig5}(c) for comparison, representing a thermalized distribution at about $k_BT=2$ keV. Although the kinetic energy spectrum becomes closer to the MB distribution for a larger system, or for $N_c$ following a log-normal distribution as found in Refs.~\cite{Zwe02,Mad04}, the resulting reaction rates are quite different. By replacing $f(v)$ in Eq.~(\ref{rate}) with the kinetic energy distribution for $N=64$ in Fig.~\ref{fig5} (c), the resulting $\langle \sigma v_{mol} \rangle$ is about $1.359\times 10^{-24}$ $\mu$m$^3$fs$^{-1}$, compared to about $3.023 \times 10^{-24}$ $\mu$m$^3$fs$^{-1}$ from the MB distribution shown in Fig.~\ref{fig5} (c). We note that the above values of the reaction rate are just for reference, since in the simulation of non-equilibrated dynamics the reaction rate depends not only on the momentum distribution both also on the coordinate information as well as its correlation with momentum. Although we can only do simulations with a limited system size and evolution time, due to the limit of the computational power in the present study, it looks unlikely that the kinetic energy spectrum may become a MB distribution in larger systems, especially for the low-energy part.

\begin{figure}[ht]
	\includegraphics[scale=0.35]{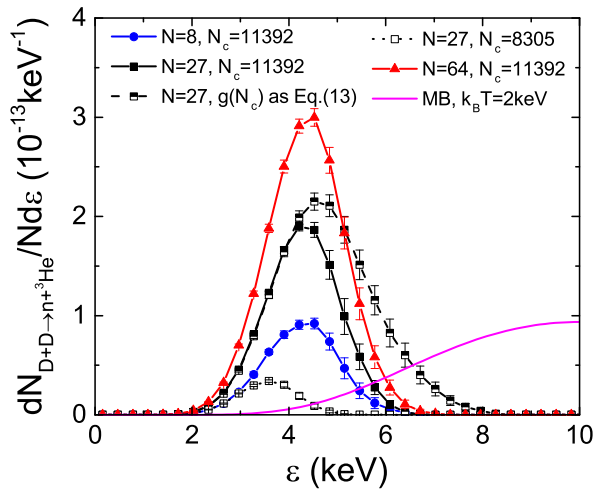}
	\caption{(Color online) Average C.M. kinetic energy distribution of D+D $\rightarrow$ n+$^3$He per cluster during the whole time period from simulating Coulomb explosions of $N=8$, 27, and 64 deuterium clusters and with different deuteron numbers $N_c$ in each cluster. The curve from a MB distribution with $k_BT=2$ keV with proper normalization is also plotted for comparison.}\label{fig6}
\end{figure}

The final kinetic energy spectrum of deuterons are generally measured experimentally, in order to extract the temperature of the system, which was compared to that extracted from the produced particles by nuclear fusion reactions in Ref.~\cite{Ban13}. However, one sees from Fig.~\ref{fig4} that neutrons are mostly produced at around $t=75$ fs, while from Fig.~\ref{fig5} the stiffness of the kinetic energy spectrum increases with time. With the kinetic energy spectrum at $t=75$ fs as in Fig.~\ref{fig5}(b), we get the reaction rate $\langle \sigma v_{mol} \rangle \approx 4.404 \times 10^{-25}$ $\mu$m$^3$fs$^{-1}$ for $N=64$, only about $1/3$ of that at $t=125$ fs. Figure~\ref{fig6} displays the average fusion reaction rate per cluster as a function of the total kinetic energy of colliding deuterons in their C.M. frame, from counting reaction number over the whole time period. Results from the MB distribution at $k_BT=2$ keV by multiplying a scaling constant is also plotted for comparison, similar to the curves in Fig.~\ref{fig1} (b) but in a linear scale, corresponding to the Gamow window. Integrating the curves over the C.M. kinetic energy $\epsilon$ leads to the corresponding neutron yield in each scenario. As is known, the peak of the distribution in Fig.~\ref{fig6} selects the colliding deuterons from their kinetic energy spectrum (also related to their spatial distribution) according to the $\epsilon$ dependence of the cross section. The distributions from simulating Coulomb explosions peak around $3-5$ keV, while that from the MB distribution at $k_BT=2$ keV with a large width peaks around 10 keV, simply because there are more deuterons at higher kinetic energies from the MB distribution compared with the case from simulations as shown in Fig.~\ref{fig5} (c). One expects that energetic deuterons inside the system at around $t=75$ fs contribute more to Fig.~\ref{fig6}, while those at the surface of the system do not contribute even if they have higher kinetic energies.

\begin{figure}[ht]
	\includegraphics[scale=0.35]{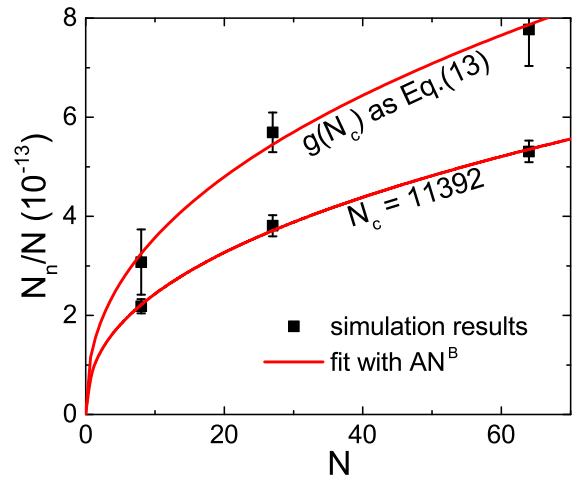}
	\caption{(Color online) Average neutron yield per cluster $N_n/N$ as a function of cluster number $N$ and fitted with an intuitive parameterized form. Results with a fixed $N_c=11392$ and a log-normal distribution of $N_c$ are compared.}\label{fig8}
\end{figure}

We now try to understand the neutron yield from systems of different sizes and extrapolate the result to even larger systems according to Eq.~(\ref{rateeq}). For clusters on regular positions in the system as in Fig.~\ref{fig3}, the system volume is $V=NV_c$, where $V_c$ is the cubic volume for cluster lattice. The life time of the system is approximated as $(NV_c)^{1/3}/ \langle v \rangle$, with $\langle v \rangle$ being the average velocity depending on the kinetic energy spectrum, consistent with the trend observed in Fig.~\ref{fig4}. In this way, the neutron yield can be approximately expressed as
\begin{equation}
N_n \sim \frac{1}{2} N V_c \rho^2 \langle \sigma v_{mol} \rangle \frac{(NV_c)^{1/3}}{ \langle v \rangle},
\end{equation}
where $\rho$ and $\langle \sigma v_{mol} \rangle$ now represent the average deuteron density and the average reaction rate during the evolution of the system. The above relation doesn't need the thermal equilibrium condition, since $\langle \sigma v_{mol} \rangle$ can be evaluated in the way similar to Fig.~\ref{fig6}. This is for the situation that clusters are on regular positions in the system, but we hope that the relation is approximately satisfied on average even for clusters at irregular initial positions. In addition, the kinetic energy spectrum of deuterons may also be different with different system sizes, as shown in Figs.~\ref{fig5} and \ref{fig6}. To account for the possible deviation from this relation, we fit the average neutron yield per cluster with the parameterized form
\begin{equation} \label{nnfit}
N_n/N = A N^B,
\end{equation}
where the coefficient $A$ is related to the average reaction rate $\langle \sigma v_{mol} \rangle$, the average cluster volume $V_c$, and the average velocity $\langle v \rangle$, etc., while the exponential constant $B$ is expected to be close to 1/3. With the simulation results of the neutron yield from $N=2^3$, $3^3$, and $4^3$, we fit the $N_n$ for different cluster numbers $N$ in Fig.~\ref{fig8}, where results for the same number of deuterons $N_c=11392$ in each cluster and those for a log-normal distribution of $N_c$, i.e., $g(N_c)$ as Eq.~(\ref{ln}), are compared. We found that such fit leads to $B\approx 0.42$ for both results, slightly larger than 1/3, which can be largely due to the stiffer kinetic energy spectrum and thus larger $\langle \sigma v_{mol} \rangle$ with increasing system size. On the other hand, the $A$ value is about $9.2\times10^{-14}$ for a fixed $N_c$ and about $13.4\times10^{-14}$ for a log-normal distribution of $N_c$. The latter is consistent with a larger average reaction rate $\langle \sigma v_{mol} \rangle$ for a log-normal distribution of $N_c$ than for a fixed $N_c$. One sees that the parameterized form of Eq.~(\ref{nnfit}) reproduces the neutron yield from systems of different sizes quite well, and can be used to extrapolate the results to larger systems. For example, with the properties of the system similar to the experimental condition in Ref.~\cite{Dit99}, our results can be extrapolated to the system as large as $100^2\pi \times 2000$ $\mu$m$^3$~\cite{Dit99}. With $V_c \approx 0.056^3$ $\mu$m$^3$, there are totally about $N=3.6 \times 10^{11}$ deuterium clusters. Based on Eq.~(\ref{nnfit}), the total neutron yield $N_n$ is about $2.6 \times 10^3$ for a fixed $N_c$ and $4.0 \times 10^3$ for a log-normal distribution of $N_c$. They are of the similar magnitude but a few times smaller compared to the neutron yield of about $10^4$ in Ref.~\cite{Dit99}. This is likely due to the smaller average $N_c$ or narrower log-normal distribution $g(N_c)$ used in the present study, the information of which is, however, not available in the experimental condition as in Ref.~\cite{Dit99}.

\section{Conclusion and Outlook}
\label{summary}

By incorporating the inelastic collision channel of D+D $\rightarrow$ n+$^3$He with the stochastic method, we have studied the nuclear fusion reactions from Coulomb explosions of deuterium clusters based on the framework of the EPOCH model. The simulations are justified by comparing the neutron yield in the box system with results from the reaction rate equation, and by comparing the Coulomb explosion results from a single cluster with available theoretical limits. We find that the kinetic energy spectrum of deuterons from the Coulomb explosion depends on the average number and the number distribution of deuterons in clusters, leading to different production powers of neutrons. On the other hand, the final kinetic energy spectrum of deuterons, which is generally used to extract the temperature of the system from experimental measurement, is different from that when neutrons are abundantly produced, and the two kinetic energy spectra correspond to significantly different reaction rates. It is found that there are less deuterons at high kinetic energies from transport simulations compared with the kinetic energy spectrum from a thermalized distribution at the same total energy. We have further investigated the dependence of the results on the cluster number, and extrapolated the neutron yield from small systems to large systems with an intuitive parameterized form. In this way, the transport approach developed in the present study has the prediction power of neutron yield in real experiments, once the detailed information, e.g., the deuteron number in each cluster, etc., is available. Our study may help to understand the nuclear fusion reactions in the non-equilibrated dynamics of Coulomb explosions induced by high-intensity laser beams.

Our study can be further generalized to incorporate other inelastic collision channels, i.e., D+D $\rightarrow$ p+T and D+$^3$He $\rightarrow$ p+$^4$He, etc. As shown in Ref.~\cite{Ban13}, the yield ratio of different particles from fusion reactions can be used to probe the temperature of the system, and such idea needs further investigation within a transport approach without the assumption of thermalization. In addition, different reaction cross sections can be introduced to study the sensitivity of the particle yield from fusion reactions to the S factor in a particular reaction, following the idea of Refs.~\cite{Bar13,Lat16}. The present study investigates the system at a relative low effective temperature of about a few keV as in Ref.~\cite{Dit99}, while the effective temperature can be 10 times higher as reached in more recent experiments~\cite{Ban13,Bar13,Lat16}, thanks to the rapid advancement of laser technology. It is of great interest to study the dynamics of the system at higher energies. Such studies are in progress.

\begin{acknowledgments}

JX was supported by the National Natural Science Foundation of China under Grant No. 11922514.

\end{acknowledgments}

\end{document}